\documentclass[aps, pre, twocolumn, amsmath, superscriptaddress,showkeys,showpacs]{revtex4-1}
\usepackage[utf8]{inputenc}
\usepackage[title]{appendix}
\usepackage{amsmath}
\usepackage{graphicx}
\usepackage{amssymb,color,tikz,multirow}
\usepackage{mathrsfs}
\usepackage{fontenc}
\usepackage{float}
\usepackage{amsthm}
\usepackage{hyperref}[colorlinks=true,linkcolor=blue,citecolor=blue]
\usepackage{algpseudocode}
\usepackage{algorithm}

\usepackage{amsfonts}

\graphicspath{ {./} }

\begin{document}

\title{A Novel Room-Based Epidemic Model: Quarantine, Testing, and Vaccination Strategies}
\author{Sourin Chatterjee}
\email{sc19ms170@iiserkol.ac.in}
\affiliation{Department of Mathematics and Statistics, Indian Institute of Science Education and Research, Kolkata, West Bengal 741246, India}
\author{Ahad N. Zehmakan}
\affiliation{School of Computing, The Australia National University, Canberra, Australia}
\author{Sujay Rastogi}
\affiliation{BITS Pilani, Department of Mathematics, Pilani Campus, India}

\begin{abstract}
Epidemic outbreaks pose significant challenges to public health and socio-economic stability, necessitating a comprehensive understanding of disease transmission dynamics and effective control strategies. This article discusses the limitations of traditional compartmental and network-based models and, inspired by the opinion formation models, introduces a room-based model that incorporates social gatherings and intuitive quarantine measures. Through simulations and analysis, we examine the impact of various model parameters, and confinement measures like quarantine and preventive measures like testing, and vaccination on disease spread. Additionally, we explore centrality-based testing and immunization strategies, demonstrating their effectiveness in reducing the spread of diseases compared to a random approach. Finally, we propose a combined strategy, that outperforms the existing strategies. It takes both global and local properties of the network structure into account, highlighting the potential for integrated control measures in epidemic management. This research not only contributes to a deeper understanding of epidemic models, but also provides insights into devising successful intervention strategies, including quarantine measures, testing methodologies, and vaccine programs to combat emerging epidemics and pandemics.
\end{abstract}

\maketitle

\section{Introduction}

Epidemic spreading has been a major concern for public health officials globally over the last two decades, with the development of numerous infectious illnesses, such as SARS, swine flu~\cite{fraser2009pandemic}, and more recently, the COVID-19 pandemic~\cite{ghosh2021reservoir,upadhyay2022combating}. These outbreaks have had a substantial influence on public health and the socio-economic situations of afflicted countries~\cite{manriquez2021spread}. Understanding the dynamics of disease transmission and executing appropriate control measures are vital for reducing the spread of infectious illnesses.

To combat such pandemics, some intervention strategies such as social distancing~\cite{qian2020covid} and contact tracing~\cite{kojaku2021effectiveness} may not be sufficient to significantly reduce the spread, especially if infected or exposed individuals are not identified accurately. Additionally, extended and stringent lockdowns may not be sustainable for a country's economic stability~\cite{shang2021effects}. An alternate solution, such as aggressive testing~\cite{upadhyay2020age, ghosh2021optimal}, may be essential to detect infected individuals efficiently. Furthermore, strategic vaccination~\cite{sartori2022comparison} (if available) can drastically prevent infection from spreading compared to random immunization.

Mathematical models involving graph theory~\cite{wang2017vaccination}, nonlinear dynamics~\cite{upadhyay2020age}, and statistics\cite{balli2021data}, ranging from stochastic to deterministic frameworks, are used to gain deeper insights into disease transmission mechanisms and analyze the efficiency of intervention strategies. Successful awareness programs, efficient contact tracking procedures, ideal vaccination techniques, and appropriate social distancing plans to postpone or eradicate the spread of a disease can be investigated using these models. 

Compartment models, such as SIR~\cite{kermack1927contribution} and SIS~\cite{hethcote1995sis}, were among the earliest attempts to imitate the disease-spreading mechanisms. In these models, the population is portioned into several groups, for example, Infected or Susceptible. Then, the states are updated following a predefined rule that mimics the spread, e.g., a Susceptible individual becomes Infected with an infection probability, as a function of the disease being modeled.

A fundamental issue inherent to compartmental models lies in their disregard for connections among individuals. In response to this issue, various disease transmission models leveraging network structures have been proposed, cf.~\cite{britton2016network}. In these models, a graph structure is used to model the interactions among individuals. Each node within the network represents an individual, whereas edges connect pairs of nodes corresponding to individuals related through friendships, professional affiliations, and other similar associations.

Although network-based models offer greater realism compared to compartment models, they frequently function under the assumption that the likelihood of disease transmission between an individual and all of their connections is uniform in nature~\cite{stegehuis2016epidemic}. Of course, in real-world settings, such uniformity does not hold true. Certain connections are stronger, increasing the possibility of large physical encounters and consequently boosting disease transmission. To address this issue, some recent works have considered the setup where each edge has a weight assigned to it, cf.~\cite{chatterjee2023effective}.

Motivated by certain opinion formation models within the area of statistical physics~\cite{galam2002minority} and inspired by real-world examples, we introduce a room-based model offering an intuitive solution to the aforementioned issue. In our model, in discrete-time rounds, the individuals are partitioned into various rooms, where the individuals in each room must form a clique in the network (that is, all the people in a room interact with each other). One can think of rooms as social gatherings, such as restaurant meetings and workplace lunches. Then, a Susceptible individual becomes Infected with a probability based on the number of Infected people in the room. A key advantage inherent to our suggested framework resides in its capacity to effectively encompass interventions such as quarantine measures in a manner that resonates intuitively.

First, we investigate how different model parameters and population sizes affect the spread of disease. We specifically look at how differences in variables like transmission rate, recovery rate, and population density impact the dynamics of the disease. Next, we shift our attention to containment measures and, in particular, the importance of quarantine in disease control. Our findings reveal that instituting quarantine measures can effectively prevent the spread of the disease. This result supports quarantine as a crucial strategy in epidemic control, as also suggested by prior studies~\cite{tang2020effectiveness,piasecki2020new}.

Furthermore, we study the effectiveness of multiple centrality-based testing and vaccination strategies. A range of measures, including degree centrality, betweenness centrality, and closeness centrality, are investigated in our study. Our results show that strategies focusing on people with a high degree of centrality regularly outperform random testing or vaccination in reducing the spread of the disease. We then propose new strategies for testing and vaccination by looking at the interactions permitted for a node in our model and find that one of the strategies named Exposure performs at a similar level of centrality-based algorithms, but this only requires local information about a node as opposed to global information in case of centrality-based strategies. Finally, we offer a Combined strategy that combines the best components of global and local information which outperforms other existing strategies.

In summary, by establishing a reliable simulation framework and bringing new, successful methodologies for vaccine allocation, and testing strategy formulation, the current work adds to the corpus of knowledge on epidemic models. Our comparison research also offers crucial insights into the efficacy of alternative approaches, paving the way for more focused and successful quarantine measures, testing methodologies, and vaccination programs in the event of emerging epidemics and pandemics.

\textbf{Roadmap.} First, we overview some further related work in more detail in Section~\ref{Work}. Then, in Section~\ref{Model} we give the precise formulation of our epidemic model. Description of the dataset and the parameters on which the experiments were performed along with several containment strategies and algorithms for testing and vaccination are discussed in Section~\ref{Experiment}. Finally, Sections ~\ref{Results} and~\ref{Discussion} present, respectively, the results of our experiments and their analysis.

\section{Related Work}\label{Work}

In this section, we discuss previous studies on various room-based models, mathematical epidemic models, and different control strategies to curb epidemics.

\subsection{Epidemic Models}

Deterministic models, such as the SIR model and its variants (SIS, SEIR, SEIS), are based on differential equations that capture transitions between compartments reflecting distinct stages of the disease. Kermack and McKendrick~\cite{kermack1927contribution} established the SIR model, which posits that the population is separated into three groups: Susceptible (S), Infectious (I), and Recovered (R). The model is characterized by its simplicity and ease of interpretation but lacks a realistic portrayal of disease spread due to its assumption of homogeneity among populations and constant contact rates. Stochastic models~\cite{pajaro2022stochastic}, in which an element of randomness is included to account for the inherent uncertainty and variability in disease spread, have been introduced to offer more realistic outcomes, especially for diseases with low prevalence or in the early stages of an outbreak when the number of infected individuals is small.

Spatial models~\cite{cliff1989spatial} employ geographic data to represent disease spread over time and place. They are especially significant for diseases like vector-borne infections since a person's geographic location has a big impact on how quickly the disease spreads. Agent-based models (ABMs)~\cite{shamil2021agent} reproduce the behavior and interactions of autonomous agents in order to analyze how their actions and interactions potentially affect the system as a whole. Each person or ``agent'' can be assigned various qualities in epidemiology that affect their susceptibility to or risk of transmitting disease. Network-based models~\cite{chatterjee2023effective} describe populations as a network of nodes (individuals) connected by edges (connections). They provide more realistic modeling of real-world interactions.

Stegehuis et al. ~\cite{stegehuis2016epidemic} conducted experiments to study dynamic processes, including bond percolation and the SIR model, on diverse network architectures. Their findings showed that the process was greatly altered by randomly shuffling the inter-community edges, but did not undergo notable changes when the edges were randomly distributed inside each community. In~\cite{upadhyay2022combating, upadhyay2020age}, an age-structured social contact matrix involving school, household, work, and others has been put to use.

\subsection{Room Based Models}

As mentioned, in our model we not only take the underlying network structure into account but also consider the social gatherings in physical spaces modeled by the distribution of the individuals into rooms. This is inspired by social gatherings in the real world and has been considered previously in the area of opinion formation and information spreading, cf.~\cite{galam2007role}.

Room-based models were first introduced in opinion formation and spreading~\cite{galam2002minority}. Using the diffusion reaction, these models explore the role of social interactions and random geometry in shaping public opinion. The social geometry of places like homes, offices, pubs, and restaurants impacts how many people assemble there. These gatherings take place sequentially across time, providing for fresh conversations and the potential for shifting perspectives. The one-person, one-argument approach is utilized to decide the outcome of debates based on local majority opinions in these models, which gives the minority no advantage. However, in case of tie reform proposal is turned down which arises from non-decision or a state of doubt. These kinds of models are now widely used in opinion dynamics~\cite{galam2007role, moussaid2013social, gartner2020threshold} and voter models~\cite{nyczka2012phase, horstmeyer2020adaptive}.

\subsection{Control Strategies}

In order to control the spread of an epidemic, several strategies can be employed, including social distancing, testing, quarantining, and vaccinations~\cite{perkins2020optimal}. Researchers use network-based epidemic models to study how the structure and evolution of networks impact disease spread and the effectiveness of interventions like vaccination and social distancing~\cite{gross2006epidemic}. It has been shown that contact tracing can be particularly effective in heterogeneous networks, as it can isolate fewer nodes while preventing more cases~\cite{kojaku2021effectiveness}.

Upadhyay et al.~\cite{upadhyay2020age} developed targeted testing strategies that focus on specific age groups to identify and isolate infected individuals, resulting in a significant reduction in the overall number of infections. In another study ~\cite{pezzutto2021smart}, authors have modeled epidemics as a stochastic process, and using the hidden Markov model they have tried to estimate the state of an individual using which they proposed a smart testing technology that outperforms traditional testing methodologies. Prioritizing testing in individuals with high degrees or betweenness can be beneficial in low testing capacities, which can reduce the total number of infections significantly~\cite{ghosh2021optimal,evans2023sociodemographic}.

Quarantine has been used long back in several infectious diseases like the Black Death, Spanish flu, Cholera, etc. and it became a tool to fight epidemics during the nineteenth century\cite{Paliga2020QuarantineAA}. Tang et al.~\cite{tang2020effectiveness} have shown that the trends of epidemics mainly depend upon quarantined cases signifying value to this particular measure. Depending on the scenario, contact tracing could have prevented from 50\% to over 90\% of cases in an epidemic, and quarantine effects are mainly limited by the fraction of undiagnosed cases~\cite{piasecki2020new}. Maximum implementation of quarantine and isolation measures during the initial phase of an epidemic has significant effects~\cite{yan2007optimal}.

Vaccination strategies have been shown to be effective in preventing disease spread and eliminating epidemics. Research has demonstrated that impulsive vaccination and achieving a certain level of vaccine coverage can control and eliminate epidemics~\cite{zeng2005complexity,aruffo2022community}. Centrality-based algorithms have been compared in random graph models, indicating that vaccination strategies based on centrality measures are more effective than random vaccination~\cite{khansari2016centrality}. Furthermore, semi-adaptive approaches that recalculate centrality measures after a fraction of nodes are vaccinated have been found to perform better than non-adaptive approaches~\cite{sartori2022comparison}. Recently, it has been shown that coupling network measures with disease parameters gives rise to a more robust vaccination strategy~\cite{chatterjee2023effective}. These findings highlight the importance of using optimal vaccination strategies to prevent and control disease outbreaks.

\section{Model Description}\label{Model}

Let $G=\left(V,E\right)$ be a simple undirected graph, where $n:=|V|$ and $m:=|E|$.  
We say that a subset of nodes $A\subseteq V$ forms (induces) a clique if for every two distinct nodes $v, u\in A$, $\{v,u\}\in E$.

Consider a graph $G=(V, E)$ which represents the network connecting the individuals in a community. Assume each node is in one of the following three states: \textit{infected} (\textit{black}), \textit{recovered} (\textit{gray}), or \textit{susceptible} (\textit{white}). More precisely, we define a \emph{coloring} to be a function $\mathcal{C}:V\rightarrow\{b,g, w\}$, where $b$, $g$, and $w$ represent \textit{black}, \textit{gray}, and \textit{white} respectively.

Assume that initially a small set of nodes is black (infected) and the rest of the nodes are white (susceptible). Then, in each discrete time round, all nodes update their color (state) according to an updating rule that we define below. Let us denote the coloring in round $t$ with $\mathcal{C}_t$.

In each round, all nodes are randomly partitioned into rooms, where the nodes in each room must form a clique in graph $G$. Then, in each room $R$, the nodes update their color in the following manner, where $0\le \beta\le 1$, $\tau\in N$ are the model's parameters:
\begin{itemize}
    \item A black node becomes gray if it has been black for the last $\tau$ rounds, and remains black otherwise. Parameter $\tau$ represents the number of days an infected node needs to recover.
    \item  We assume that a node that has been infected and recovered will not become infected again and that is why a gray node remains gray forever.
    \item The parameter $\beta$ models \textit{transmission rate} of the virus. The larger values of $\beta$ indicate that the virus can spread easily. We assume an infected (black) node can make a susceptible (white) node, in the same room, infected with probability $\beta$. Therefore, the probability that a white node becomes black in a room with $b$ black nodes is $1-(1-\beta)^b$. 
\end{itemize}

To complete the formulation of the model, we need to explain how exactly the random partition of the nodes into rooms takes place. Let $\mathcal{K}$ be the set of all node sets that form a clique in $G$. (We allow cliques of size $1$ to account for nodes who are less ``social'' and attend gatherings less often.) We start with all nodes and all cliques. Then, we pick a clique among all available cliques uniformly at random and assign the nodes on that clique to a room. Now, we eliminate the nodes from our node set and all cliques which include at least one of the eliminated nodes, and repeat the process for the updated node set and clique set. This way, in the end, each node is assigned to exactly one room.

\begin{figure*}[ht]
\centering
  \includegraphics[width=1\linewidth]{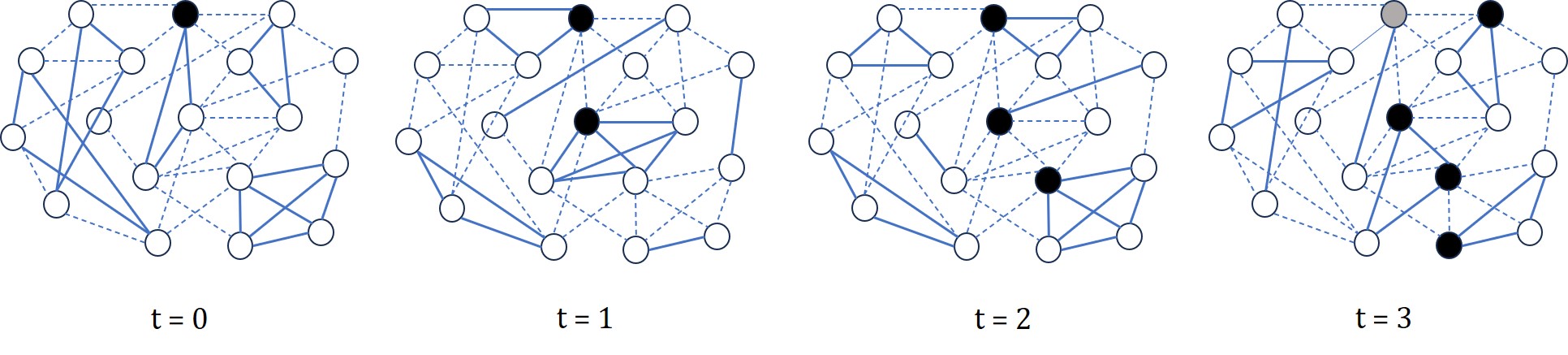}
  \caption{Visual representation of epidemic spread model over the course of time. Susceptible (white), infected (black), and recovered (grey) are shown in the figure. In each time step, nodes that are partitioned into the same room have been connected via solid lines, and the remaining connections are shown using dotted lines. Here we have $\tau=3$.}
  \label{fig:model}
\end{figure*}

In FIG. \ref{fig:model}, we provide an example of our disease-spreading model. Starting with 18 nodes (out of which one is infected (black)), we partition the nodes into rooms, and then the disease spreads within each room independently.

\section{Experimental Setup and Strategies}\label{Experiment}

In this section, we will discuss the model parameters and real-world graphs that have been put into use to carry out experiments. Furthermore, we will talk about the synthetic graph models being used to mimic real-world networks and different parameter variations being used to study the epidemic process. Finally, how several containment strategies are implemented and algorithms to prioritize testing and vaccination are discussed. 

\subsection{Real-world and Synthetic Networks}

For our experiments, we utilize both real-world graph data and synthetic graph models. For real-world networks, we rely on publically available data from SNAP~\cite{leskovec2009community,leskovec2012learning}. Finding cliques for a very large graph is computationally expensive and due to limited computational capacity we choose 800 to 900 nodes. Moreover, it is worth noting that due to privacy reasons a large part of the network is not available. In particular, we ran our simulations on the Facebook dataset, the Twitter dataset, and the Slashdot dataset of which some description is given below.

\begin{itemize}
    \item \textbf{Facebook} is a platform for social networking. An undirected graph contains a representation of every user-to-user link in the Facebook network. An edge between two nodes $v$ and $u$ indicates the friendship between two individuals. The network has $900$ nodes and $9976$ edges. 
    \item \textbf{Twitter} (known now as X) is a social networking website where users upload and engage with messages known as ``tweets'' on this platform.  If user $v$ follows user $u$ then there is an edge from node $v$ to node $u$. We have converted this graph to an undirected one using the rule: an edge $e$ exists between $u$ and $v$ if there is an edge from node $v$ to node $u$ or node $u$ to node $v$. It has $800$ nodes and $9521$ edges. 
    \item \textbf{Slashdot} is a website for technological news that is submitted by users and later reviewed by editors. A feature on this website allows users to tag other users as friends or enemies. This network contains links between users. Though it was a directed network we have converted it to an undirected one using a similar procedure as Twitter. It has $900$ nodes and $7524$ edges. 
\end{itemize}

It is important to note that the graph data obtained from online social platforms may not be an exact match for our particular use case, since it is possible for individuals to be connected on social media without any physical interaction. However, we have chosen to use this data for various reasons.
First of all, real-world social networks, whether online or physical, share certain graph characteristics such as a small diameter, scale-free degree distribution, and a large clustering coefficient, as noted in previous studies. Therefore, although the graph data we use may not perfectly match physical interactions, it still possesses these desirable properties, as supported by our experiments using synthetic graph data. Secondly, there is a large amount of graph data available from online social networks, which allows us to conduct a more extensive and comprehensive set of experiments than would be possible with physical interaction data.  Lastly, in our model, two nodes that share more friends are more likely to be assigned to a room. This is aligned with the observation that two friends who live closely in a geographic sense are more likely to meet in person because people with large overlaps between their friends usually live in similar locations.

Furthermore, various synthetic random graph models have been put out to resemble real-world social networks, cf.~\cite{costa2007f}.
These models are typically designed to have basic characteristics that are consistently seen in actual networks, like small diameter and power-law degree distribution.

We use a random graph model called the Hyperbolic Random Graph (HRG), which produces complex networks with hyperbolic geometry. The HRG model contains nodes that are embedded in hyperbolic space~\cite{gugelmann2012random}. According to the HRG model, nodes are pulled close to one another based on their geometrical closeness in the hyperbolic space.

In our experiments, the HRGs are generated in such a manner that the number of nodes and edges match with the experimented real-world networks (namely Facebook, Twitter, and Slashdot graphs from above) using the Networkit Python Package~\cite{staudt2016networkit}.

An HRG is generated with 4 parameters: the number of nodes ($n$), the average degree of a node ($K$), temperature ($T$), and the exponent of the power-law degree distribution ($\gamma$). It is known that clustering is maximized at $T=0$, minimized at $T=\infty$, and undergoes a phase transition at $T=1$. As a result, for $T<1$, the graph displays clustering behavior, however for $T>1$, the clustering goes to 0~\cite{krioukov2009curvature}. According to a study by Krioukov et al.,~\cite{krioukov2009curvature}, the internet graph has a temperature of $T=0.6$ when it is embedded in hyperbolic geometry. Thus, we also set $T=0.6$ in our setup. In addition, we use $b = 2.5$ because it has been empirically shown that in social networks, $2 \leq b \leq 3$~\cite{albert2002statistical}. We chose the average degree similar to our tested real-world networks, namely Facebook, Twitter, and Slashdot.

\subsection{Parameter Variations}

The default values for the model's parameters are $\beta=0.5$ and $\tau=5$. Furthermore, we only focus on cliques of size $1$ to $6$ for complexity reasons. It is worth emphasizing that there is nothing special about these choices. However, they seem to be suitable choices to capture some real-world examples such as COVID-19. We also study the behavior of the model for different values of these parameters.

In order to study how the parameters: recovery time ($\tau$) and transmission rate ($\beta$) influence the disease dynamics, keeping one parameter fixed other parameters are varied, and their effects are studied here. To investigate the effect of $\tau$, we set $\beta = 0.5$ and $\tau$ is varied in the range [1, 2, 4, 5, 8, 16]. Similarly, we keep $\tau = 5$ and vary $\beta$ in the range [0.01, 0.16, 0.2, 0.32, 0.4, 0.64, 0.8] to study its effects. To observe how the size of the graph affects the lasting of the epidemics, HRG models are created having different population sizes from 32 to 1024 and how many days it takes to obtain a completely disease-free population is studied. 

For all the experiments, we assume that initially $10$ randomly selected nodes are infectious. These numbers are chosen to ensure that the virus almost surely spreads to a large part of the network; otherwise, it is not very meaningful to apply containment measures such as testing or vaccination strategy. Furthermore, for all graphs, we run our experiments $10$ times and use the average value. All the experiments have been implemented using Python 3 and NetworkX library~\cite{hagberg2008exploring}.

\subsection{Containment Measures and Startegies}

As discussed previously, testing, quarantining, and vaccinating are the key strategies in order to mitigate disease. Now, in this scenario, we ask the question of how effective any mass quarantine or lockdown strategy can be depending upon a portion of people not following it or how late it has been imposed. Moreover, we also try to come up with effective testing and vaccination strategies in order to minimize the number of infected individuals.

\paragraph{Quarantine.} While imposing quarantine, we do not permit the nodes to change their rooms (cliques), except for a portion of the population who are not following the rules. On each day, the fraction that is not following quarantine is chosen at random.

\paragraph{Vaccination.} In the case of vaccination, only susceptible nodes can be vaccinated and once vaccinated they develop an ever-lasting immunity to the disease, i.e. they can not develop the infection.

\paragraph{Testing.} For testing purposes, we only consider nodes that are susceptible, infected, and half of the recovered nodes. We assume that the other half of the nodes self-report being recovered, and hence they are not tested. Also, once a node tests positive for the disease, it is kept in isolation for $\tau$ rounds (that is, until it recovers). 

\paragraph{Vaccination/Testing Strategies.} We assume that we are given the budget to vaccinate or test up to $\alpha$ percentage of the population (or equivalently $k=\lfloor \alpha n\rfloor$ individuals) and the ultimate goal is to minimize the total number of infected nodes during the process.

\paragraph{Centrality-based Algorithms.} A natural approach is to use standard algorithms used to select the most ``influential'' nodes in a graph such as the highest degree, highest closeness, and highest betweenness, cf.~\cite{petrizzelli2022beyond}. 

\paragraph{Proposed Algorithms.} One natural parameter to measure the influence of a node $v$ in our model is to compute the number of cliques which include $v$. Thus, we propose an algorithm that vaccinates/tests nodes with the highest clique inclusion. Another suitable choice is to compute the number of nodes that are in at least one clique with $v$. This is essentially the number of nodes that can be infected by $v$. We also introduce a Combined algorithm that combines a centrality-based algorithm and one of the proposed algorithms. 

In our experiments, the $\alpha$ percentage of nodes with the highest score, according to some scoring mechanism, are vaccinated or tested among the available nodes (see Algorithm~\ref{alg}). Thus, below, we simply need to define what score function is used in each strategy. For example, in the Degree algorithm, the score of a node is its degree. We should emphasize that none of our algorithms assumes any knowledge of the state of the network (i.e., which nodes are infected/susceptible). In other words, all algorithms are ``source-agnostic''. 

\begin{algorithm}[H]
\label{alg}
\caption{Find Nodes to Vaccinate/ Test}
\label{algo:get_top_alpha_fraction}
\begin{algorithmic}[1]
\State Calculate \textsc{Score}$(v)$ for each node $v$. 
\State Sort all nodes according to \textsc{Score} in descending order.
\State Find the list $L$ of $\lfloor \alpha n\rfloor$ nodes with the highest \textsc{Score}.
\State \textbf{return} $L$.

\end{algorithmic}
\end{algorithm}

\subsection{Algorithms for Testing and Vaccination}

All the $9$ vaccination and testing strategies, that have been put to the test to minimize the number of infected nodes are listed below along with their description:

\begin{enumerate}
    \item \textbf{Random}: This algorithm randomly chooses nodes for vaccination (i.e., assigns a random and unique score to each node). 
    \item \textbf{Degree}: It measures the number of nodes to which that node is connected: \[d\left(v\right):=|N\left(v\right)|.\]

    (For example, in this algorithm, \textsc{Score}$(v)=d(v)$ and the $\lfloor \alpha n\rfloor$ nodes with the highest degree are vaccinated.)

    \item \textbf{Betweenness}: Betweenness centrality is a measure of a node's importance in a network based on the premise that a node is important if it lies on many shortest paths between other nodes in the network. The number of overlaps with the shortest paths between pairs of nodes is then used to determine how central a node is:
\[b(v) := \sum_{s \neq v \neq u} \frac{\sigma_{su}(v)}{\sigma_{su}}\]
where $\sigma_{su}$ the total number of shortest paths from node $s$ to node $u$ and $\sigma_{su}(v)$ is the number of those paths that pass through $v$.

    \item \textbf{Closeness}: Based on the notion that a node is important if it is close to other nodes in the network, closeness centrality is a measure of a node's relevance in a network. The inverse of the sum of the shortest distances between a node and every other node in the network is then used to establish a node's centrality. Formally, the closeness centrality of a node is given by:
\[c(v) := \frac{n-1}{\sum_{u\neq v} d(u,v)}\]
where $d(u,v)$ is the length of the shortest path between $u$ and $v$, disregarding the weights.
    \item \textbf{Eigenvector}: A node's significance in a network can be determined by looking at how it is connected to the other significant nodes in the network. In other words, the sum of the centralities of a node's neighbors defines its centrality. More precisely, given the adjacency matrix $A$ of a graph $G=(V,E)$, we define
    \[x(v) := \frac{1}{\lambda} \sum_{u\in V} A_{u,v} x(u)\]
    Then, $X=[x(v_1),x(v_2),...,x(v_n)]^T$ is the solution of the equation $AX = \lambda X$ and the $i$-th component of $X$ will give eigenvector centrality score of node $v_i$.

    \item \textbf{Pagerank}: Pagerank centrality is a measurement of a node's significance or centrality inside a network. It calculates the probability that a random walker will arrive at a specific node after moving through the network by following links. A node is viewed as more influential and central to the network the greater its pagerank centrality. Given a graph $G=(V,E)$, pagerank centrality of a node is given by,
    \[
    p(v) = (1 - \kappa) + \kappa \sum_{u \in N(v)} \frac{p(u)}{d(u)}
    \]
    Here, $\kappa \in (0,1)$ is a damping factor whose value is taken to be 0.85. As we are considering an undirected graph, we used the degree of $u$, otherwise, we need to consider the out-degree of $u$.
    \item \textbf{Cliques}: This algorithm counts the number of cliques a particular node is part of.

    \item \textbf{Exposure}: A node's influence can be understood by counting the number of different nodes it is interacting with via a room-assigning mechanism. So, it essentially counts the number of different nodes in cliques the particular node is part of. 

    \item \textbf{Combined}: In this combined algorithm, we take into account the Exposure and Betweeness of a node, and nodes are scored accordingly, giving both of them equal weights.

\end{enumerate}

\section{Results}\label{Results}
We provide our experimental findings in two subsections. First, we discuss how changing the model parameters can impact the outcome of the epidemic, and then we present our results on how different containment strategies can curb the disease spread.

\subsection{Effects of Model Parameters}\label{Parameter Results}

We investigate the impact of each of the three model parameters of recovery period ($\tau$), transmission rate ($\beta$), and size of population ($n$).  

\subsubsection{Effect of Recovery Period}

\begin{figure*}[ht]
\centering
  \includegraphics[width=1\linewidth]{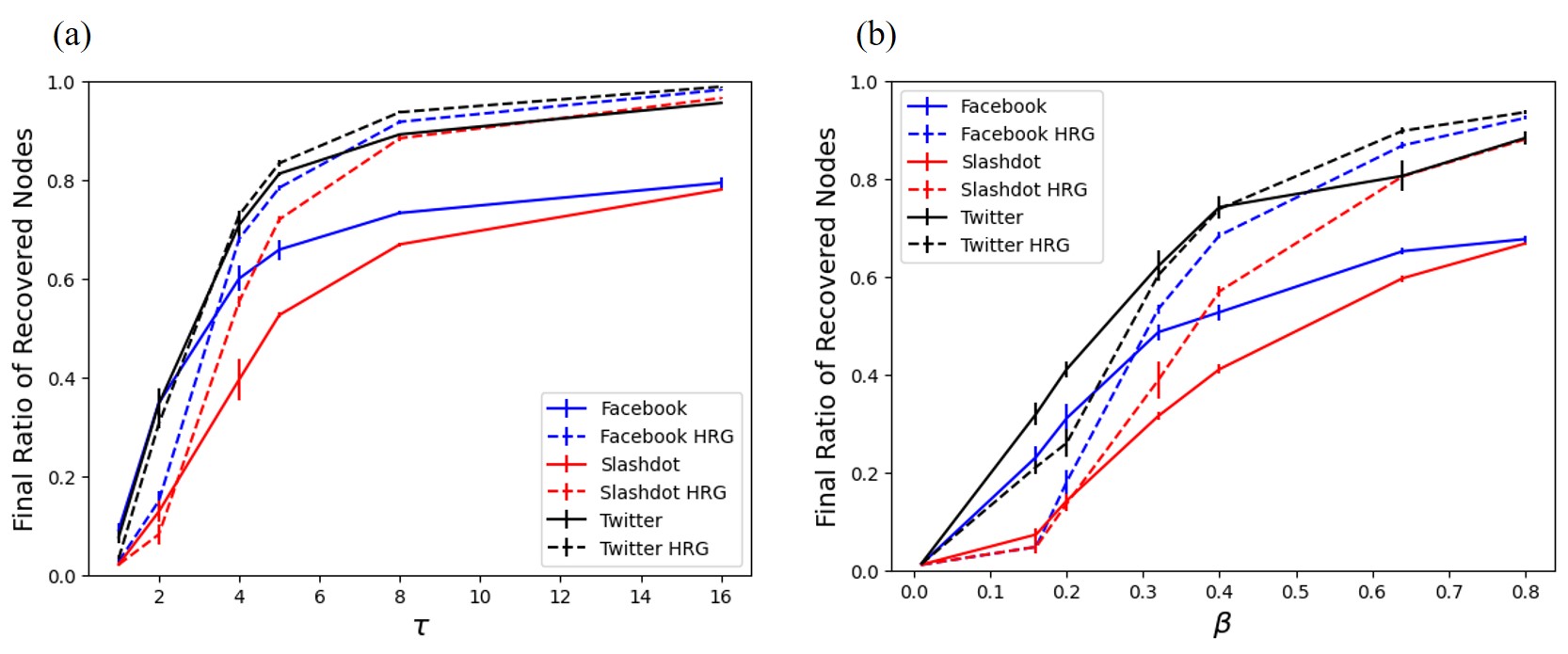}
  \caption{Effect of (a) Recovery Period and (b) Transmission Rate on Recovered Nodes in the Network. The error bar represents the standard deviation of the results.}
  \label{fig:comp_tau_beta}
\end{figure*}

As we increase the recovery period, the number of infected nodes and in turn the final number of recovered nodes should be higher which can be seen in FIG. \ref{fig:comp_tau_beta}. Moreover, we observe a linear growth with respect to increasing $\tau$ for smaller values of $\tau$. However, a threshold behavior appears when $\tau=8$, which means changing the recovery period does not affect the number of infected nodes significantly. This can be explained via the point that for $\tau=7$ already around 70\% of nodes become infected and thus the effect of increasing $\tau$ starts becoming less and less significant.

Furthermore, we observe that Facebook HRG (the HRG graph generated with parameters similar to the Facebook graph) exhibits a behavior reasonably similar to the Facebook graph. This is an indication that HRG is a suitable model. Similar behavior is observed for Twitter and Slashdot.

\subsubsection{Effect of Transmission Rate}

Increasing transmission rate causes higher infection spreading resulting in a higher final number of recovered nodes. We observe in FIG. \ref{fig:comp_tau_beta} that after $\beta=0.4$, increasing $\beta$ does not affect the total number of recovered nodes drastically, but below that value, we see a very rapid growth in recovered nodes. This is again because at $\beta=0.4$ already almost 60\% of nodes become infected (and then recover); thus, the effect of increasing $\beta$ after that becomes less and less significant.

Again, the HRG counterparts exhibit relatively similar behavior.

\subsubsection{Effect of Population Size on Epidemic Duration}

Note that the process eventually reaches a fixed configuration where all nodes are either susceptible or recovered. We call the number of rounds the process needs to reach such configuration \textit{convergence time}. We are interested to know how the convergence time of the process changes as the number of nodes $n$ grows. 

\begin{figure*}[ht]
\centering
  \includegraphics[width=0.5\linewidth]{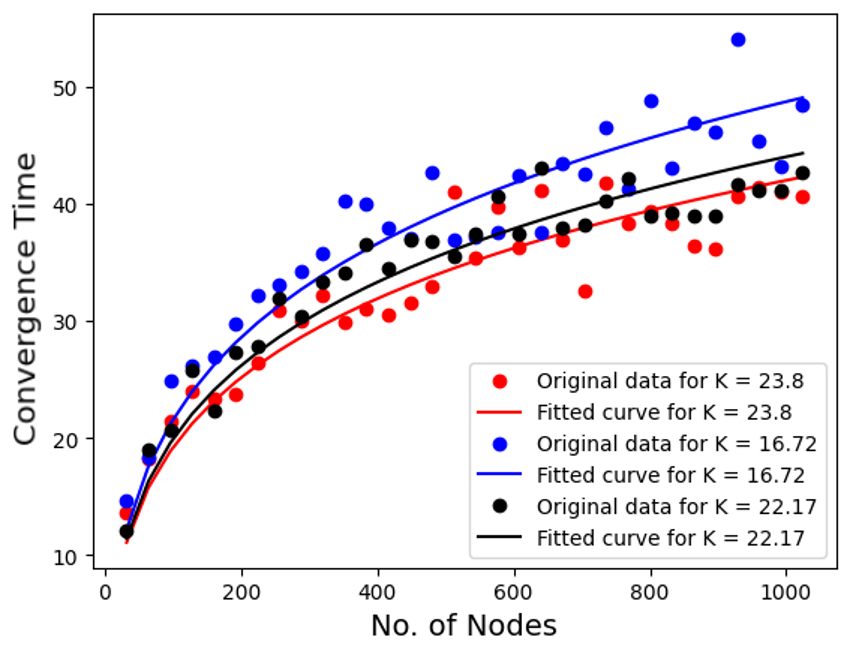}
  \caption{Convergence time for different numbers of nodes in the HRG Graph. Here K represents the average degree of a graph. }
  \label{fig:rounds_nodes}
\end{figure*}

Since real-world social networks are of a given size, they are not suitable for these experiments. Instead, we have measured the convergence time of the process for the synthetic HRG graph for various values of $n$, and the results are demonstrated in FIG. \ref{fig:rounds_nodes}.

We observe that as we increase the average degree of the network the convergence time is improved. Furthermore, as we increase the size of the network, the convergence time becomes larger, as one might expect. We find using fitting that the asymptotic growth seems to correspond to a poly-logarithmic function. So, if we denote the number of nodes by $n$, and the convergence time by $t$, then the following relation seems to be true,
\[t \propto (\log(n))^\eta\]

where $\eta$ is a coefficient that depends upon the average degree of the graph and if the average degree of the graph is higher, we find a lower value of $\eta$.  

The correlation with $\eta$ can be explained by the fact that a larger degree increases the exposure and interactions between the nodes and consequently results in a faster spread and convergence.

Since we only consider cliques/rooms of constant size, the number of susceptible nodes can increase at most by a constant factor, regardless of the random choices. Based on our experiments, only a small fraction of nodes are susceptible at the end of the process. This explains why the process would need at least logarithmically many rounds to converge.

On the other hand, when a graph has strong expansion properties (such as a complete graph or binomial random graph), in each round an infected node has a constant probability of infecting a susceptible node. Thus, the convergence time would be logarithmic. While the real-world social networks and HRGs are not strong expanders, they enjoy some level of expansion. This should explain the conjectured poly-logarithmic bound from above.

\subsection{Effects of Containment Strategy}\label{Strategy Results}

In this subsection, we first discuss how variation in quarantine imposition and the percentage of nodes following can alter the outcome of the disease spread. Then, we provide a comprehensive comparison of various vaccination and testing strategies.

\subsubsection{Quarantine}

\begin{figure*}[ht]
\centering
  \includegraphics[width=1\linewidth]{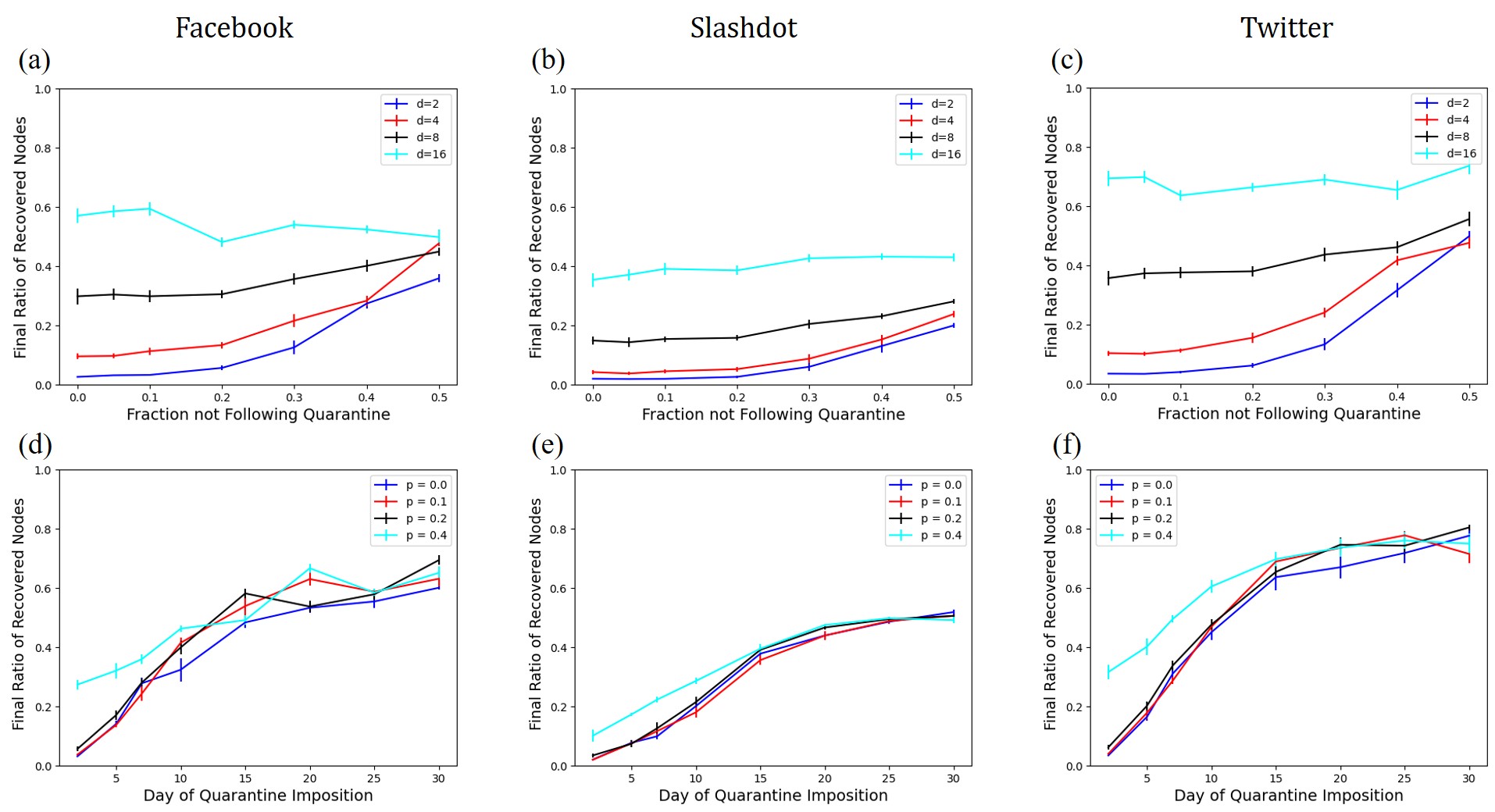}
  \caption{The effect of quarantine on the number of infected nodes. The top row includes how the final ratio of recovered nodes changes with the fraction of nodes not following quarantine rules in the presence of different quarantine imposition times denoted by $d$. The bottom row shows the final ratio of recovered nodes changes with the day of quarantine imposition for a different fraction of people not following quarantine denoted by $p$. The left, middle, and right column depicts the results for Facebook, Slashdot, and Twitter respectively. }
  \label{fig:reco_all}
\end{figure*}

Using Quarantine, we want to measure how a delayed response can affect the outcome of an epidemic. As expected, using all three graphs (Facebook, Slashdot, and Twitter), we find that if the quarantine is imposed early and followed by most people the number of infected nodes (consequently, the final number of recovered nodes) is significantly smaller as seen in FIG. \ref{fig:reco_all}.

Interesting results come from the interplay between these two parameters. When the quarantine is imposed after 16 days, what fraction of people are following the quarantine does not really matter. It is important to note that this threshold strongly depends on the size of the population and the model's parameters. Furthermore, if a significant fraction of the population does not follow the quarantine, then the infection spreads to a large fraction of the population, regardless of how early the quarantine is imposed.

\begin{figure*}[ht]
\centering
  \includegraphics[width=1\linewidth]{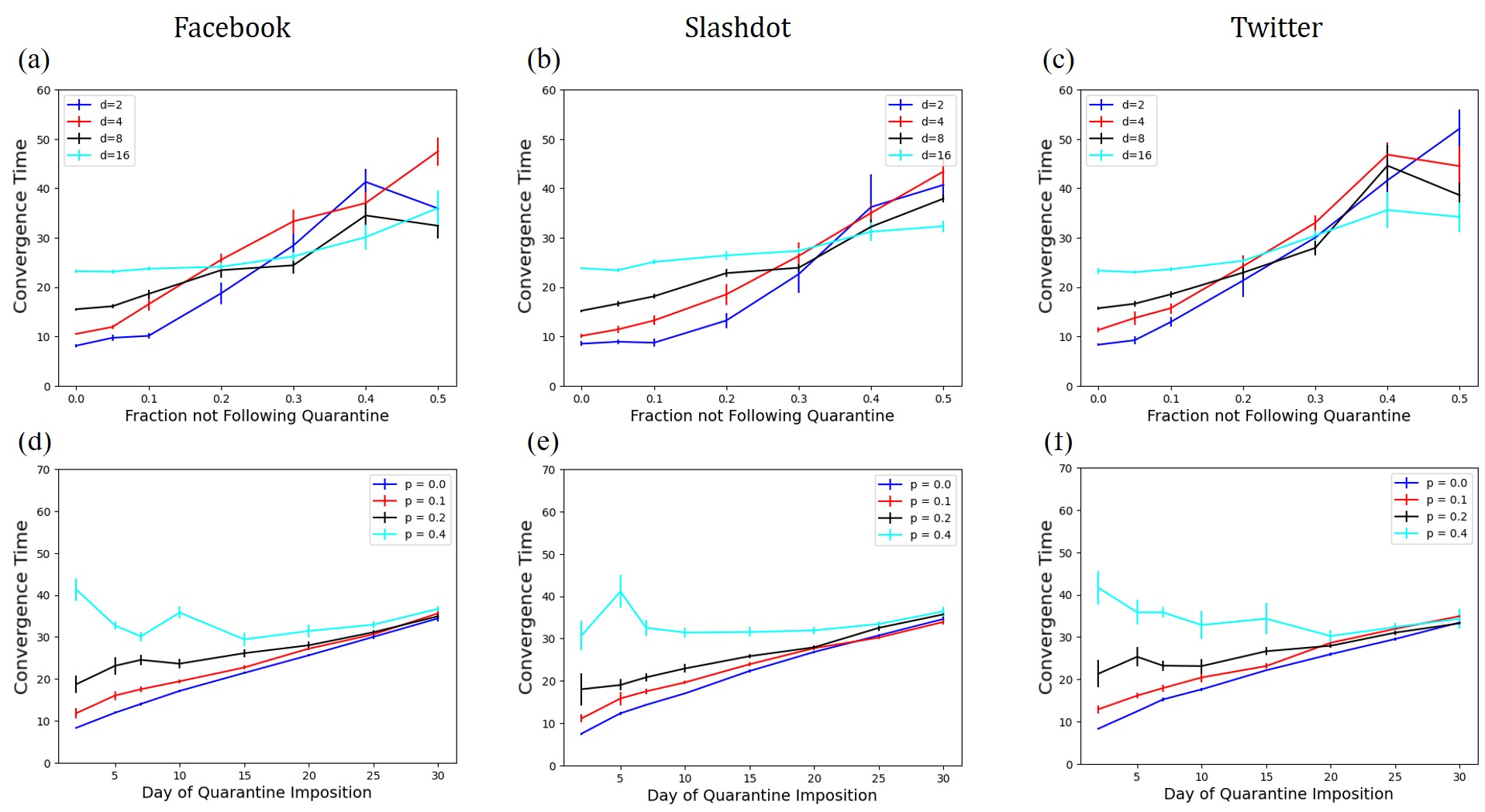}
  \caption{The convergence time while quarantine is imposed. The top row shows how, in the presence of various quarantine imposition times represented by $d$, the proportion of nodes not adhering to quarantine regulations alters the length of the epidemic. The bottom row demonstrates how the length of the epidemic varies depending on the day a quarantine is imposed for a particular fraction of people, denoted by $p$. The results for Facebook, Slashdot, and Twitter are shown in the left, middle, and right columns, respectively.}
  \label{fig:days_all}
\end{figure*}

We also investigated the effect of these two parameters, the percentage of nodes violating quarantine and the day the quarantine is imposed, on the convergence time of the process (that is, the number of days it takes for the process to stabilize). The results are depicted in FIG. \ref{fig:days_all}.

We observe that for a fixed imposition day, when the fraction of nodes not following quarantine increases, the convergence time increases too. One explanation is that this increases the chances of the spread continuing and results in a larger convergence time. However, when the fraction is too large, for example, 0.5, then it becomes smaller since the spread becomes faster.

For a fixed fraction of nodes not following quarantine measures, varying the day the quarantine is imposed results in a larger convergence time. A natural explanation is that this delay adds to the convergence time and the infection has spread to a larger body of network once quarantine is imposed, which makes the process take even longer. When the fraction not following is too large, e.g. 0.4, a different behavior is observed. One explanation is that if the quarantine is imposed later, the infection has been spread and the process is already almost in stabilization.

\subsubsection{Testing}

\begin{figure*}[ht]
\centering
  \includegraphics[width=1\linewidth]{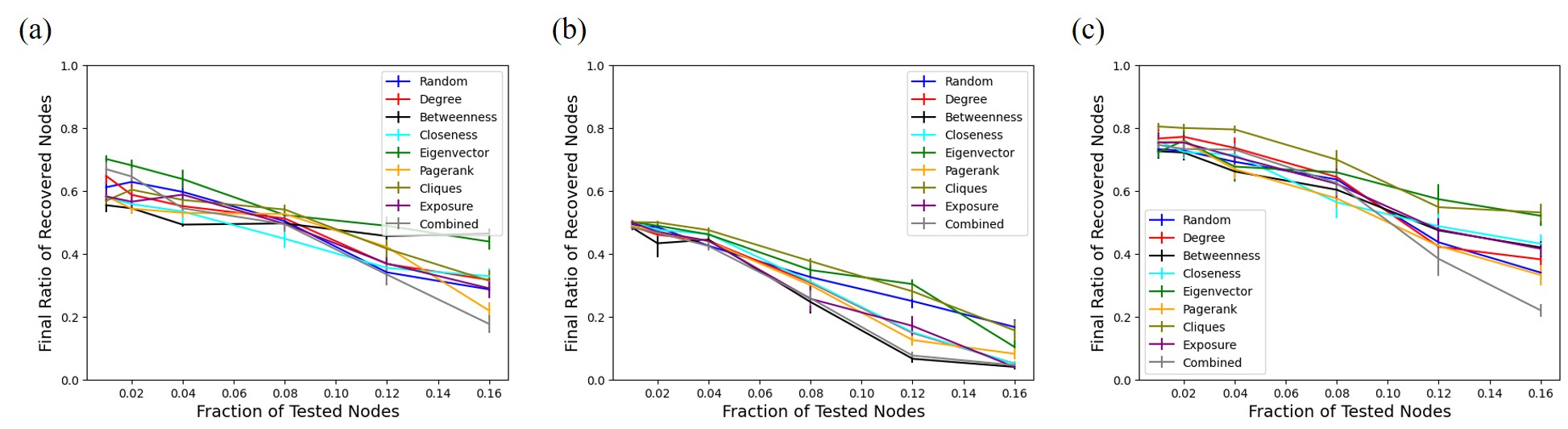}
  \caption{The effect of testing with different strategies on (a) Facebook (b) Twitter (c) Slashdot.}
  \label{fig:testing}
\end{figure*}

As we increase the fraction of tested nodes we find that the total ratio of recovered nodes goes down which can be observed in FIG. \ref{fig:testing}. Algorithms like Eigenvector and Cliques perform very badly to reduce the number of infected nodes. Random algorithms perform fairly across all scenarios. On the other hand, Betweenness, Closeness, Exposure, and Combined algorithms performed very well across graphs and different fractions of testing. In Facebook and Twitter graphs we find that our Combined algorithm outperforms others significantly in the higher fraction of testing. But, in the case of Slashdot Betweenness is the best performer though results from the Combined algorithm are comparable, and when 0.16 fraction of nodes are tested all better-performing algorithms give almost similar results. 

\subsubsection{Vaccination}

\begin{figure*}[ht]
\centering
  \includegraphics[width=1\linewidth]{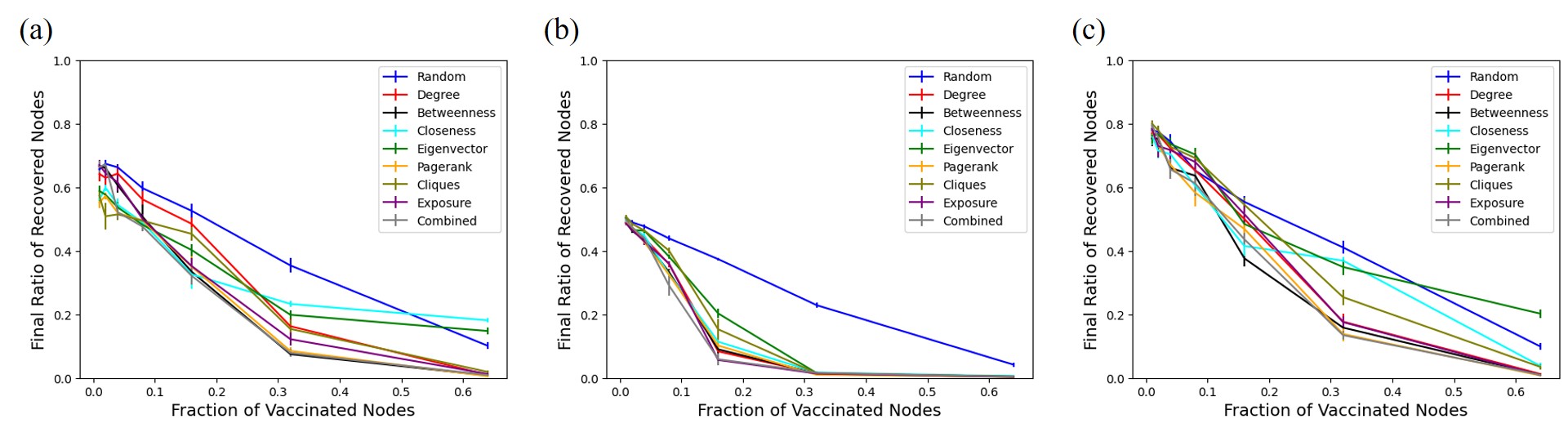}
  \caption{The effect of different vaccination strategies on (a) Facebook (b) Twitter (c) Slashdot.}
  \label{fig:vaccination}
\end{figure*}

In FIG. \ref{fig:vaccination}, we can observe that the ratio of recovered nodes goes down with the increasing vaccination coverage. Random vaccination performs worst in all scenarios. Generally, Betweenness, Exposure, Pagerank, and Combined algorithms performed well, and their results are very similar when 60\% of the nodes are vaccinated. Also, we find in the case of Slashdot, that using 32\% vaccination with any strategies except Random gives us the best possible results. It is also important to note Combined algorithm does not improve the result by a huge margin but it constantly performs well in almost every scenario and more so when vaccine coverage is around 32\%.

\section{Discussion}\label{Discussion}

We have provided a new model of epidemic spreading in graphs which can be very helpful in understanding effective control strategies for epidemics. We evaluated how introducing quarantine can help in containing disease progression and along with that we have tried several testing and vaccination methods to find out the best strategy among them. Our proposed algorithms proved to be more successful at increasing vaccination or testing percentages.  However, betweenness centrality generally performed very well, suggesting that it is vital to consider centrality measures for designing testing or vaccination strategies. One of our contributions is designing a Combined algorithm that outperforms existing vaccination and testing methods or at least performs equally depending upon the percentage of the available resources. In most situations, this Combined algorithm performed better than other approaches, highlighting the potential advantages of merging several methodologies into a single, cohesive strategy. Our findings are in accordance with a prior study that discovered centrality indicators to be useful in vaccination tactics~\cite{kitsak2010identification}. It is crucial to stress that these conclusions are robust because numerous trials were conducted, minimizing the impact of randomness. Furthermore, our testing and vaccination plan is independent of the knowledge of the nodes that were initially infected. 

While considering the variation of model parameters, we found that among three different networks, Twitter has the highest ratio of infected nodes and Slashdot has the lowest at all the parameter values. Now, this can be explained by the fact that Twitter has the highest average degree and Slashdot has the lowest. Though we discover that, HRGs slightly overestimate the infection spread, we can state that the results in the synthetic graphs are fairly comparable to that of the real networks which approve that the HRG graph models real-world networks up to a very good extent~\cite{gugelmann2012random}. 

Note that the severity or impact of an epidemic is not directly linked with the epidemic duration, as a short epidemic can occur with people getting infected very fast or very less people getting infected at all. Hence, we find that when there is a large fraction of nodes are not following the quarantine and quarantine is imposed late, the disease gets over quickly in comparison to cases where quarantine is imposed early.

One interesting observation is while that Random vaccination performs poorly, the effect of Random testing is acceptable. This is perhaps because in the vaccination a set of randomly chosen nodes are vaccinated once, while in testing every time we pick a random set of nodes to test. The random testing in some sense gives an edge to this strategy in comparison to others, since they behave deterministically.


Though we have shown successful vaccination and testing strategies, there are a few limitations of the study. Firstly, due to a lack of information regarding the disease parameters, we have considered all populations having homogeneous infection rates and recovery periods but chances of other diseases, age, and sex can have an effect on these parameters. Secondly, our study focused on static networks, but real-world contact networks are typically dynamic and constantly changing. Furthermore, there might be distinctions between social networks and the real-world connections of those who propagated the epidemic~\cite{zhang2014comparison}. Thirdly, we have not taken into account re-infection or illness even after vaccination, which may be the case in real-world situations~\cite{sciscent2021covid}. Fourthly, due to the lack of huge computational resources this study was done with a small number of nodes but if the conclusion regarding vaccination and testing policies still holds true needs to be checked. Next, in a real-world population, the network structure of connection may not be fully known hence identifying key nodes according to the centrality metrics can be a challenge, but methods like exposure only take local information, and recently it has been shown that demographic information like 
age, gender, marital status, educational attainment, and household size can be used as a proxy for centrality measures~\cite{evans2023sociodemographic}. Finally, the practicality of deploying these algorithms in genuine scenarios like logistical obstacles, and public acceptability are not taken into account here. 

The primary objective of this study is to investigate the efficacy of several testing, vaccination, and quarantine techniques within a novel room-based SIR-type model framework, assuming no possibility of re-infection. However, it is worth noting that future research could explore alternative models, such as the SIS model, which allows for re-infection among individuals. Such models would be valuable in replicating the transmission dynamics of diseases where immunity is not long-lasting or permanent, such as the Common Cold or Influenza~\cite{hethcote1989three}. Additionally, while our current analysis assumes a scenario of vaccinating the entire population simultaneously, it is important to acknowledge that real-world vaccination programs often adopt step-by-step approaches with semi-adaptive measures~\cite{sartori2022comparison}. Investigating the impact of such vaccination strategies in future studies would provide valuable insights.

Finally, we can say that by integrating local and global information into the network effective testing and vaccination strategies can be developed, and along with that quarantine strategies can also be incorporated keeping in mind the economic variable. Integration of social and economic parameters into these kinds of studies will be a key step in the future where proper epidemic control policies can be designed. The usage of available computational models will be crucial for designing public health policies in order to curb future pandemics.  

\section*{Authors' Contribution}
\textbf{Sourin Chatterjee}: Conceptualization, Methodology, Software, Validation, Formal analysis, Investigation, Visualization, Writing. \textbf{Ahad N. Zehmakan}: Conceptualization, Methodology, Validation, Formal analysis, Writing. \textbf{Sujay Rastogi}: Conceptualization, Methodology, Software, Visualization, Writing. 

\section*{Declaration of Competing Interest}
The authors declare that they have no known competing financial interests or personal relationships that could have appeared to influence the work reported in this paper.

\bibliographystyle{apsrev4-1}

\end{document}